\documentclass[%
reprint,
superscriptaddress,
 amsmath,amssymb,
 aps
]{revtex4-1}

\usepackage{amsmath}
\usepackage{algorithm}
\usepackage[noend]{algpseudocode}

\makeatletter
\def\BState{\State\hskip-\ALG@thistlm}
\makeatother

\usepackage{graphicx}% Include figure files
\usepackage{dcolumn}% Align table columns on decimal point
\usepackage{bm}% bold math
\usepackage[flushleft]{threeparttable} % table footnote

\usepackage{dsfont}
\usepackage{color,psfrag}
\usepackage{mathtools,amscd}

\usepackage{soul}
\usepackage{xcolor} % colored tables
\definecolor{purple}{rgb}{0.5,0.0,0.5}
\definecolor{mypink1}{RGB}{219, 48, 122}
\definecolor{mypink2}{cmyk}{0, 0.7808, 0.4429, 0.1412}
\definecolor{mygray}{gray}{0.6}
\definecolor{mygreen}{HTML}{0CA172}
\definecolor{myblue}{HTML}{20517D}

\newcommand\here[1]{\fcolorbox{red}{red}{\rule{0pt}{6pt}\rule{6pt}{0pt}}\quad}
\newcommand\missing[1]{\fcolorbox{blue}{blue}{\rule{0pt}{6pt}\rule{6pt}{0pt}}\quad}

\usepackage{ifpdf}
    \ifpdf
\usepackage{tikz}
\usetikzlibrary{arrows,chains,matrix,positioning,scopes, fit, calc}
\usetikzlibrary{cd}
\usepackage{chemfig}
\fi
\usepackage{hyperref}

\hypersetup{
	colorlinks,
	linkcolor=blue,
	citecolor=blue, 
	filecolor=black,
	urlcolor=blue}

\begin{document}

\preprint{APS/123-QED}

%%%%%%%%%%%%%%%%%%%%%%%%%%%%%%%%%%%%%%%%%%%%%%%%%%%%%%%%%%%%%%%%%%%
%%%%%%%% =========================== T I T L E ============================%%%%%%%%%%%%%%%%
%%%%%%%%%%%%%%%%%%%%%%%%%%%%%%%%%%%%%%%%%%%%%%%%%%%%%%%%%%%%%%%%%%%

% Machine Learning enabled battery electrodes characterization: The Janus anode case of study for sodium ions batteries % Force line breaks with \\
%
\title{
Characterizing High-Capacity Janus Aminobenzene–Graphene Anode for Sodium-Ion Batteries with Machine Learning}
\author{Claudia Islas-Vargas}
\affiliation{Instituto de F\'isica, 
Universidad Nacional Aut\'onoma de M\'exico, Cd. de M\'exico C.P. 04510, Mexico}
\author{L. Ricardo Montoya}
\affiliation{Instituto de F\'isica, 
Universidad Nacional Aut\'onoma de M\'exico, Cd. de M\'exico C.P. 04510, Mexico}
\author{Carlos Vital}
\affiliation{Instituto de F\'isica, 
Universidad Nacional Aut\'onoma de M\'exico, Cd. de M\'exico C.P. 04510, Mexico}
\author{Oliver T.\ Unke}
\affiliation{%
Google DeepMind
}
\author{Klaus-Robert M\"uller}%
\email{klaus-robert.mueller@tu-berlin.de}
\affiliation{%
 Machine Learning Group, Technische Universit\"at Berlin, 10587 Berlin, Germany
}
\affiliation{%
BIFOLD -- Berlin Institute for the Foundations of Learning and Data, Berlin, Germany
}%
\affiliation{%
Google DeepMind
}
\affiliation{%
Department of Artificial Intelligence, Korea University, Anam-dong, Seongbuk-gu, Seoul 02841, Korea
}%
\affiliation{%
Max Planck Institute for Informatics, Stuhlsatzenhausweg, 66123 Saarbr\"ucken, Germany
}
\author{Huziel E. Sauceda}
\email{huziel.sauceda@fisica.unam.mx}
\affiliation{Instituto de F\'isica, 
Universidad Nacional Aut\'onoma de M\'exico, Cd. de M\'exico C.P. 04510, Mexico}
\date{\today}% It is always \today, today,

\begin{abstract}
Sodium-ion batteries require anodes that combine high capacity, low operating voltage, fast Na-ion transport, and mechanical stability, which conventional anodes struggle to deliver.
Here, we use the SpookyNet machine-learning force field (MLFF) together with all-electron density-functional theory calculations to characterize Na storage in aminobenzene-functionalized Janus graphene (Na$_x$AB) at room-temperature. Simulations across state of charge reveal a three-stage storage mechanism—site-specific adsorption at aminobenzene groups and Na$_n$@AB$_m$ structure formation, followed by interlayer gallery filling—contrasting the multi-stage pore-, graphite-interlayer-, and defect-controlled behavior in hard carbon. 
This leads to an OCV profile with an extended low-voltage plateau of 0.15 V vs. Na/Na$^{+}$, an estimated gravimetric capacity of $\sim$400 mAh g$^{-1}$, negligible volume change, and Na diffusivities of $\sim10^{-6}$ cm$^{2}$ s$^{-1}$, two to three orders of magnitude higher than in hard carbon. Our results establish Janus aminobenzene–graphene as a promising, structurally defined high-capacity Na-ion anode and illustrate the power of MLFF-based simulations for characterizing electrode materials.
\end{abstract}

\pacs{Valid PACS appear here}% PACS, the Physics and Astronomy
                            
\maketitle

%\tableofcontents

%%%%%%%%%%%%%%%%%%%%%%%%%%%%%%%%%%%%%%%%%%%%%%%%%%%%%%%%%%%%%%%%%%%
%%%%%%%%%% ========================== INTRODUCTION ========================== %%%%%%%%%%%%%
%%%%%%%%%%%%%%%%%%%%%%%%%%%%%%%%%%%%%%%%%%%%%%%%%%%%%%%%%%%%%%%%%%%

\section{Introduction} \label{intro}

Sodium-ion batteries (SIBs) are increasingly viewed as a sustainable and cost-effective complement to lithium-ion systems due to the abundance and broad geographic distribution of sodium resources on Earth.~\cite{Chayambuka, batteriesLIBvsSIB_review, Usiskin, Passerini, Scrosati2014_grapehene_aergy_storage}
Yet, progress remains hindered by the poor affinity of graphite for Na$^+$, a consequence of unfavorable intercalation thermodynamics due to a large ionic radius, high atomic mass, and slow kinetics.\cite{2016_Yuanyue_PNAS,Scrosati2014_grapehene_aergy_storage}
This limitation has led to the exploration of alternative carbon-based anodes capable of stabilizing Na$^+$ intercalation. 
Among these, hard carbons have emerged as leading candidates, where sodium storage involves a combination of defect adsorption, interlayer insertion, and the formation of quasi-metallic sodium clusters within confined pores.\cite{Dahn2000_High_capacity_materials_for_SIBs, Liu2024_insituClusters, Yu2025_HC_Na_clusters_AIMD} 
Despite their success, the intrinsic structural heterogeneity and low ion-diffusivity values of hard carbons complicate systematic optimization and limit the ability to tailor Na-host interactions by design.\cite{Chen2022_HC_hybridMechanism, Liu2024_insituClusters}
Interestingly, recent studies show that two-dimensional carbonaceous materials can exhibit sodium-storage motifs analogous to those of hard carbons, but arising from surface chemistry rather than structural disorder.\cite{Tiong2025_HighPerformance_SIB_graphene,  Advancements_graphite_electrodes_LIB_SIB, Tian2024_Janus_Cobalt_LIB_SIB_experimental, Samori2022_Janus2D_functionalized_materials}
Graphene derivatives retain high electrical conductivity and mechanical stability, while their surface chemistry can be systematically modified to engineer favorable sodium-hosting environments. Approaches such as heteroatom doping, covalent attachment of organic groups, or asymmetric Janus functionalization can simultaneously expand interlayer spacing and introduce high-affinity adsorption sites, thereby improving Na$^{+}$ accommodation relative to pristine graphene.~\cite{Advancements_graphite_electrodes_LIB_SIB, Tian2024_Janus_Cobalt_LIB_SIB_experimental} Among these approaches, Janus 2D materials are especially attractive because asymmetric functionalization generates intrinsic dipolar fields and heterogeneous binding domains that enhance Na-ion adsorption, facilitate diffusion, and improve structural stability during cycling. \cite{Liu2025_Janus_hexapentagraphene_KIB_anode, Peimanirad2025_ABG_Na_DFT, janusGraphene2021, Samori2022_Janus2D_functionalized_materials}.
In this context, functionalizing graphene with aminobenzene groups creates a polar structure which tailors both the electrostatic potential and the interlayer environment, making it a promising material for high-performance SIB anodes.
However, accurately characterizing the electrochemical behavior of functionalized Janus 2D materials under realistic operating conditions presents a significant challenge.~\cite{Samori2022_Janus2D_functionalized_materials, Tian2024_Janus_Cobalt_LIB_SIB_experimental} Their polar nature, heterogeneous surface chemistry, and Na-ion adsorption/desorption processes lead to a complex interplay between electrostatics, dispersion, and charge transfer effects.~\cite{Peimanirad2025_ABG_Na_DFT, Liu2025_Janus_hexapentagraphene_KIB_anode}
Moreover, finite-temperature effects, including fluctuations in interlayer spacing and dynamic rearrangements of active sites, cannot be captured by static zero-temperature electronic structure calculations.~\cite{2016_Yuanyue_PNAS, Scrosati2014_grapehene_aergy_storage} These considerations highlight the need for simulation frameworks capable of treating quantum-mechanical interactions with chemical accuracy while accessing the nanosecond timescales relevant for ion transport.~\cite{Unke.MLFF.ChemRev2020,ChemRev_Deringer2021,2025_MLFF-chap_Wiley} 
% MLFFs
Machine-learning force fields (MLFFs) trained on \textit{ab initio} data provide such a capability, enabling predictive modelling of ion interactions and structural dynamics without sacrificing electronic-structure fidelity.~\cite{2021_HDNN4,2022_bigdml,2024_FFLUX,2022_NequiP,2022_MACE,2023_Allegro,2024_SO3krates,2021_SOAP-vdW}
In the last decade, MLFFs have become a cornerstone on molecular and materials simulations, providing highly accurate and efficient models enabling predictive simulations in chemistry~\cite{Rupp_CM_PRL2012,ANI_ChemSci2017,Noe.MLFF.ARPC2020,von2018quantum,von2020exploring,QML-Book,musil2021physics,keith2021combining,kabylda2025molecular,2025_DeltaML_Cazares}, physics~\cite{Veit.Methane.JCTC19,Cheng.HighPress-Hydro.Nat2020,sauceda2021NatCommun,Deringer.SOAP-Si.Nat2021,VDOS_Shapeev2020},  biology~\cite{GAO.MLFF-biology.Patterns2020,Noe.MLFF-biology.CPSB2020,unke2024biomolecular}, and materials science~\cite{Goedecker_IonSolids_PRB2015,Shapeev_Moment_MLSciT2021,Artrith.JChemPhys148.2018,Bartok_SOAP-Si_PRX2018}.

\begin{figure}[htp]
    \centering
    \includegraphics[width=1.0\columnwidth]{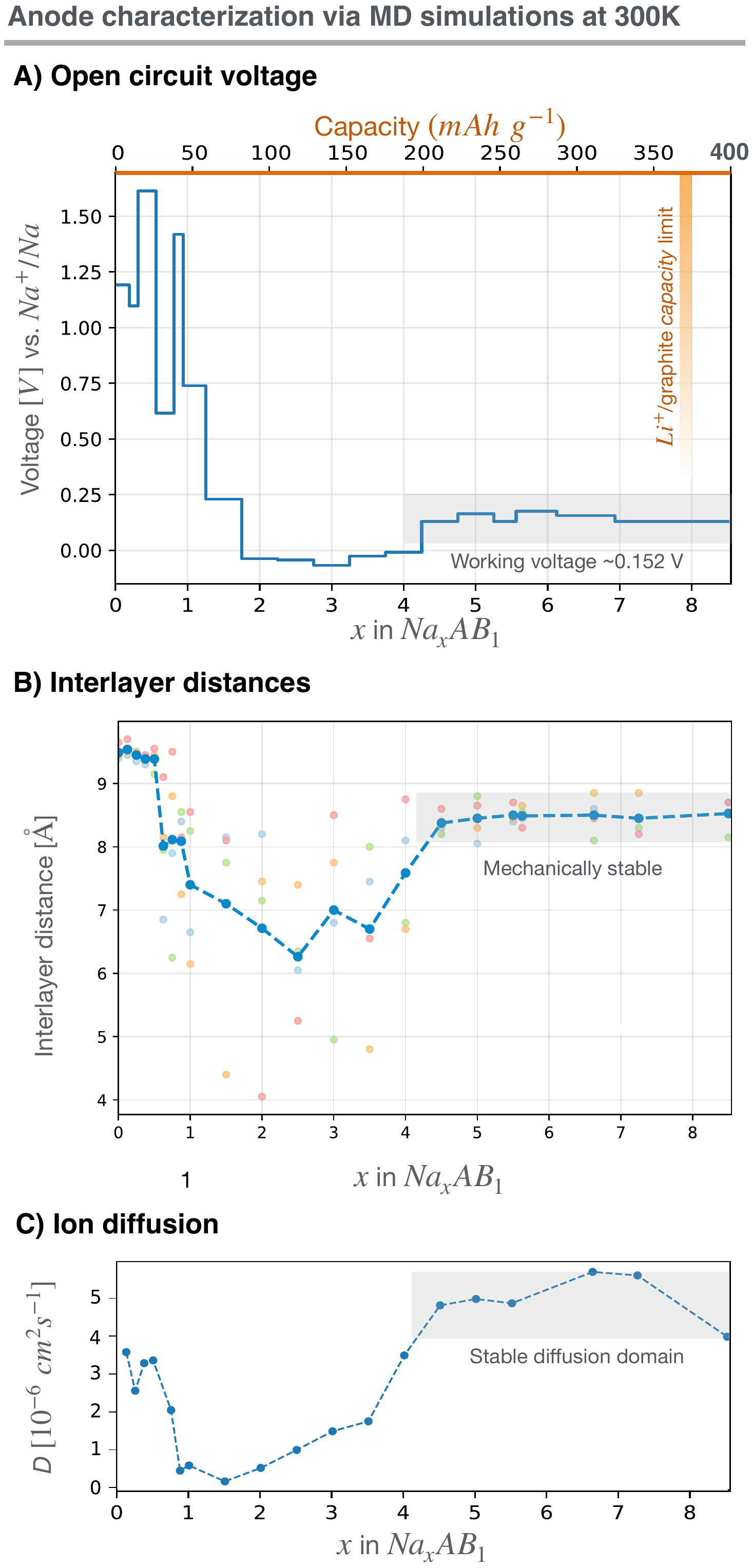}
    \caption{\textbf{Anode characterization as a function of state of charge.} \textbf{A)} Open-circuit voltage profile extracted from Machine-learning Molecular Dynamics at 300 K. \textbf{B)} Average interlayer distance between graphene layers as Na content increases. The individual dots display the specific intra-plane distances. \textbf{C)} Sodium-ion diffusion coefficient computed from the mean-squared displacement. 
    The grey-boxes highlight the plateaus for the voltage, volume and ion diffusivity reached over the high-loading stable domain: The \textit{working state-of-charge} regime.}
    \label{fig:AnodeChar}
\end{figure}

In this article, we present a case study with application to a candidate SIB anode material called Janus graphene.~\cite{janusGraphene2021} This material is a challenging quantum system that displays a wide variety of interactions, such as electrostatic (e.g. \mbox{Na$^+$--Na$^+$} interactions, \mbox{Na$^+$--NH$_2^-$}, and cation-$\pi$ interactions), dispersion (e.g., benzene--benzene, graphene--graphene and benzene--graphene), as well as covalent and metallic bonding (i.e., formation of metal Na clusters). At finite temperatures, the Janus graphene material is a prominent dynamical system, showing ubiquitous electrochemical activity, charge exchange, and strong fluxional motion at all frequency scales.
Given such diverse types of interactions, we have used the SpookyNet MLFF~\cite{unke2021spookynet} due to its robustness in describing local and delocalized interactions trained on Density Functional Theory (DFT) calculation~\cite{FHIaims2009} with the nonlocal Many-Body Dispersion correction~\cite{2020_MBDnl}.
We performed molecular dynamics simulations at various states-of-charge, obtaining room-temperature physical and electrochemical properties, insights into its internal ion dynamics, and a careful characterization of the anode's open circuit voltage (OCV), mechanical stability, and diffusion coefficient (Figure~\ref{fig:AnodeChar}). See Methods section in Supporting Information for further details.
The Janus graphene material displayed solid anode capabilities with a stable desirable voltage ($\approx$0.15 V), a negligible volume change, and a consistent high ion-diffusivity ($\approx$ 5$\times$10$^{-6}$cm$^2$/s) in the denominated \textit{working state-of-charge} regime. 
%

%%%%%%%%%%%%%%%%%%%%%%%%%%%%%%%%%%%%%%%%%%%%%%%%%%%%%%%%%%%%%%%
%%%%%%%%%% ========================== RESULTS ========================== %%%%%%%%%%%%%
%%%%%%%%%%%%%%%%%%%%%%%%%%%%%%%%%%%%%%%%%%%%%%%%%%%%%%%%%%%%%%%
\section{Results and discussion}\label{Results}

\begin{figure*}[htp]
\centering
\includegraphics[width=0.95\textwidth]{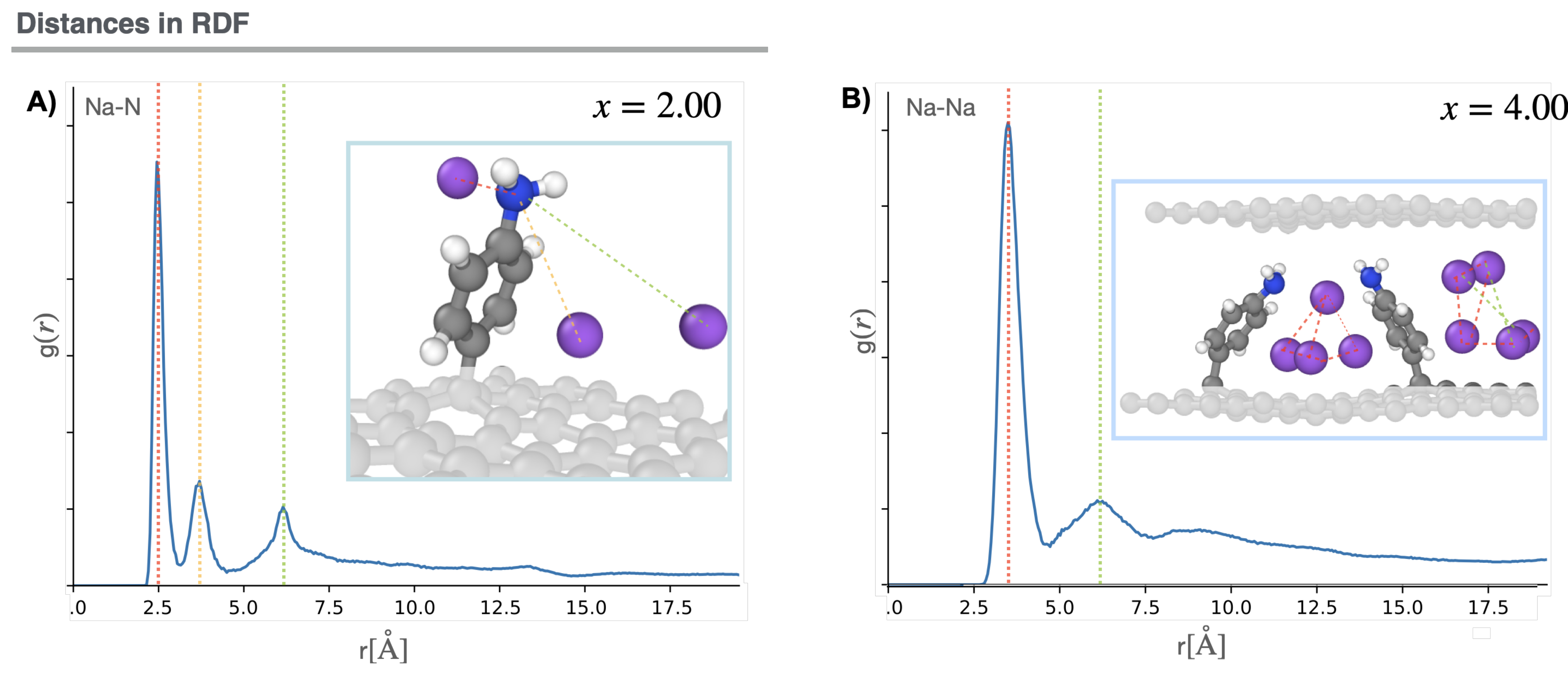}
\caption{\textbf{Radial Distribution for Na--N and Na--Na interactions in Na$_{x}$AB.} \textbf{A)} Radial distribution function $g(r)$ for Na–N pairs at $x = 2.0$. The first peak at approximately 2.5 \AA{} (red dashed line) corresponds to Na directly adsorbed on the \mbox{--NH$_{2}$} groups, while the second peak near 4.0 \AA{} (yellow dashed line) arises from Na atoms positioned above the aromatic rings. A broader peak at 7.0 \AA{} (green dashed line) reflects more weakly correlated Na ions located in adjacent interlayer environments. 
\textbf{B)} Na--Na radial distribution function at $x=4.0$. The first peak at 3.0 \AA{} (red dashed line) indicates short-range Na--Na correlations characteristic of Na-cluster formation. A broader second peak at 6.5 \AA{} (green dashed line) corresponds to Na atoms in higher coordination environments, including those situated on the graphene layers. 
Dashed lines in the schematic illustrate the spatial arrangement associated with these coordination distances.}
\label{fig:Fig_gr}
\end{figure*}

\begin{figure*}[htp]
    \centering
    \includegraphics[width=1.0\textwidth]{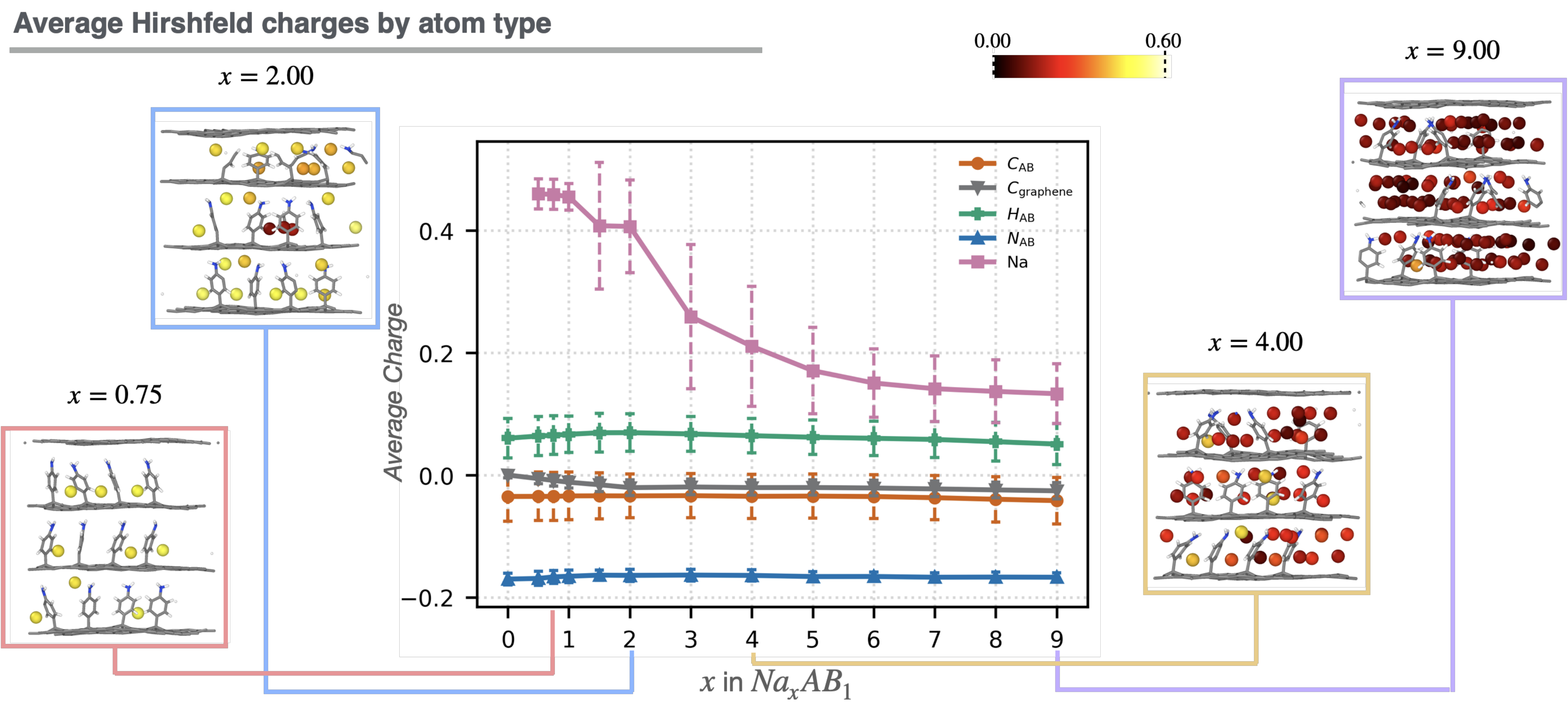}
    \caption{\textbf{Average Hirshfeld charges for each atomic species as a function of sodium content ${x}$.} C$_\mathrm{AB}$, H$_\mathrm{AB}$, and N$_\mathrm{AB}$ denote atoms in the aminobenzene groups. Values represent averages over ten MD configurations per composition; error bars indicate the standard deviation. Snapshots embedded in the plot show representative configurations, with Na atoms colored according to their Hirshfeld charge (color scale: black=0, white = 0.6).}
    \label{fig:Fig_Average_H_Charges_vs_Na_x}
\end{figure*}

\begin{figure}[htp]
    \centering
    \includegraphics[width=1.0\columnwidth]
    {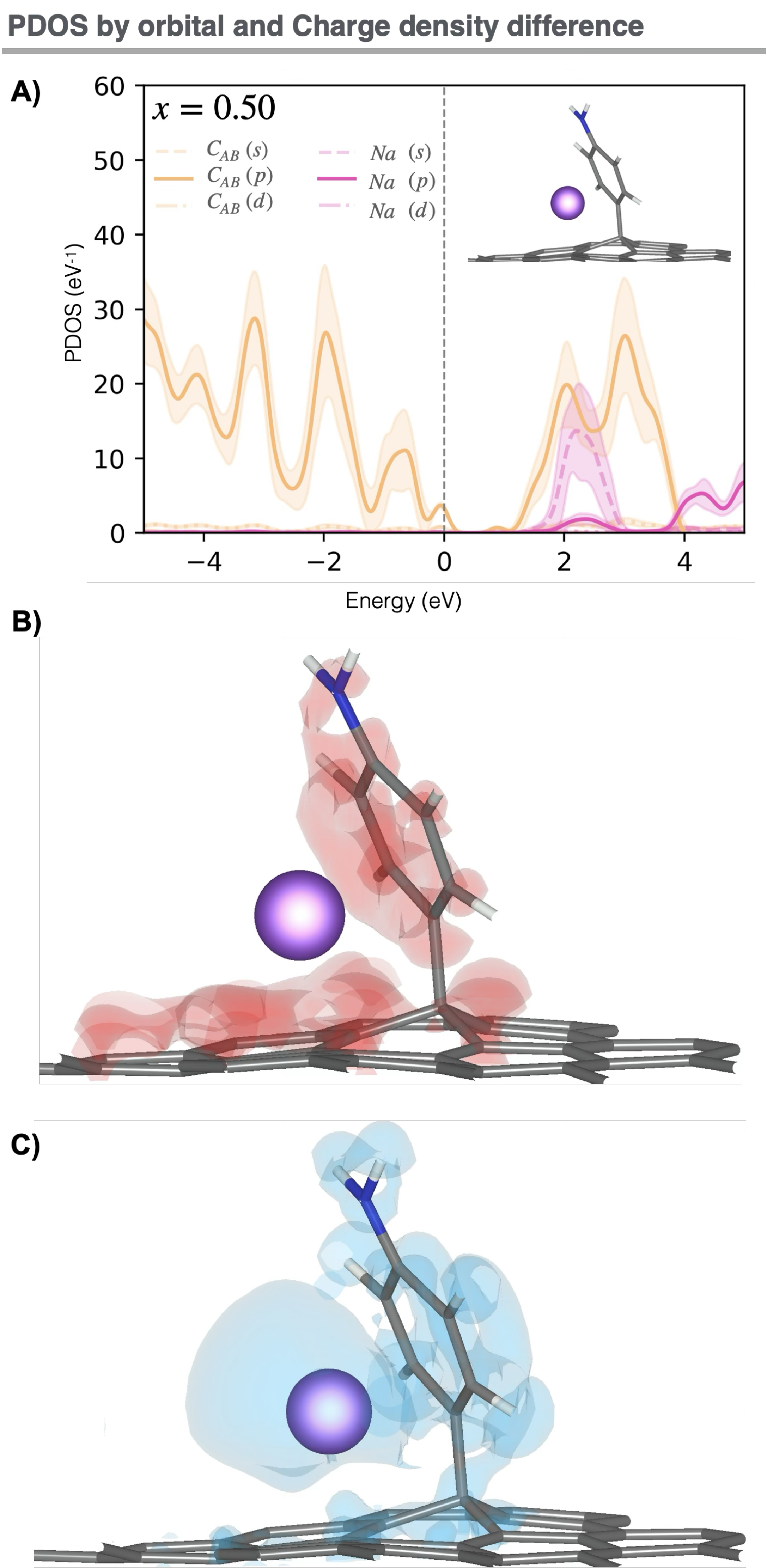}
    \caption{\textbf{Orbital-resolved electronic structure and charge-density difference for $x =0.5$.} \textbf{A)} Projected density of states (PDOS) averaged over ten single-point calculations. Shaded regions indicate the standard deviation across sampled configurations. \textbf{B)} and \textbf{C)} Charge-density difference ($\Delta{\rho{}}$) isosurfaces for a representative configuration at $x =0.5$, plotted at ±0.004 e \AA$^{-3}$. Red denotes electron accumulation and blue denotes electron depletion. The $\Delta{\rho{}}$ reveals depletion around the Na atom and extended accumulation over the benzene ring, characteristic of a cation–$\pi{}$ interaction. }
    \label{fig:delta_rho_x0.5}
\end{figure}

\textit{Na adsorption and interlayer response to increasing ion concentration.}
The $Na^{+}/Janus$ (Figure S1 A) system can be synthesized with high precision in the number of layers, where, experimentally, Sun \textit{et al.}~\cite{janusGraphene2021} built anodes with up to 8 layers. 
Here, we focus on the 4-layer system, which presents the most challenging dynamical behavior.
To elucidate how asymmetric functionalization governs sodium accommodation in Janus aminobenzene–graphene (Na$_{x}$AB), we first examined the structural response of the anode across states of charge.
ML-driven simulations at room temperature reveal that Na insertion induces a restructuring of the Na$_{x}$AB framework. Figure \ref{fig:AnodeChar} summarizes the electrochemical and structural response of Janus aminobenzene–graphene as a function of sodium concentration. 
At low Na content ($x<0.5$), individual Na$^{+}$ preferentially adsorb near the amino-benzene groups, occupying sites in proximity to the nitrogen sites and adjacent $\pi$ systems (see insets in Figure~\ref{fig:Fig_gr}), while the interlayer spacing remains close to the pristine value (9.5 \AA~at $x=0$).
As the Na concentration increases, the adsorbed ions interact strongly with the amino-benzene groups, inducing a progressive tilting of the amino-benzene pillars driven by a coupled interaction graphene$-$Na$^+-$benzene-ring (Figures~\ref{fig:Fig_gr}B and~\ref{fig:delta_rho_x0.5}B).
These structural rearrangements draw the graphene layers closer together, reducing the average interlayer distance (Figure \ref{fig:AnodeChar}B).
In particular, sodium concentrations in the range $0.5<x\le 4$ present a large availability and diversity of adsorption sites, whose occupancy is highly fluctuating due to thermal excitations inducing frequent structural reconfigurations.
At intermediate and high sodium concentrations ($x\ge 3$), Figure \ref{fig:AnodeChar}B displays the emergence of mechanical stability, which starts with the occurrence of aminobenzene pyramidal-like structures stabilized by a small sodium cluster: Na$_n$@AB$_m$ structures (Figures \ref{fig:Fig_Average_H_Charges_vs_Na_x} at $x=2$, \ref{fig:Fig_gr}B-inset, S7 and S8 at $x=3$).
Consequently, the restricted fluxionality of the saturated aminobenzene molecules prompts interlayer stability and creates channels to host additional Na ions.
Overall, the cooperative interplay between the Janus aminobenzene–graphene backbone and the dynamical formation of sodium clusters stabilizes the anode framework and suppresses volumetric expansion to negligible levels, thereby conferring mechanical robustness while maintaining a high sodium storage capacity (Figure \ref{fig:AnodeChar}B).

\textit{Charge redistribution and Na adsorption sites.} 
The Janus aminobenzene–graphene framework exhibits a wide variety of local chemical environments and ion-adsorption sites, each of which withdraws a different amount of electronic charge from the adsorbed sodium atoms.
 Therefore, we examine the charge redistribution associated with Na adsorption using Hirshfeld analysis 
to understand the microscopic origins of the material's mechanical and electronic response to an increasing state of charge.
Figure~\ref{fig:Fig_Average_H_Charges_vs_Na_x} shows that, at a low state of charge, the Na atoms exhibit a highly ionic character, consistent with localized ion-host interactions (see also Figure S2 for $x\lesssim 2$). As Na concentration increases, the average charge of sodium atoms decreases steadily; nevertheless, the charge variance increases in the interval $1\lesssim x\lesssim 5$, indicating ion environment heterogeneity. This result is consistent with the structural variability displayed in Figure~\ref{fig:AnodeChar}B).
At high loading ($x>5$), both the mean charge and its variance stabilize at lower values, signaling partial charge delocalization and metallic-like behavior, in agreement with cluster formation visible in the structural snapshots of Figures~\ref{fig:Fig_Average_H_Charges_vs_Na_x} and S2. 

Previous DFT studies at equilibrium configurations report that the Na$^+\cdots$NH$_2-$ configuration is $\sim$4 kcal/mol more stable that Na$^+\cdots$aromatic-ring (Bz).~\cite{janusGraphene2021, Peimanirad2025_ABG_Na_DFT} 
Contrasting this energetic hierarchy, our finite-temperature analysis shows that, statistically, the Na$^+\cdots$Bz adsorption site, driven by the cation-$\pi$ interaction~\cite{Dougherty2025_CationPi_ChemRev, Petrushenko2020_CationPi_Inorganic}, is substantially more populated than the amino site (Figures~\ref{fig:Fig_Average_H_Charges_vs_Na_x} and S2).
This preference arises because the aromatic $\pi$-cloud provides a broader (i.e., larger cross-section) and more polarizable electron density that stabilizes Na$^{+}$ through strong electrostatic induction contributions, with reported binding energies of $\approx{}$ 20 kcal/mol and only minor dispersion involvement.~\cite{Dougherty2025_CationPi_ChemRev, Petrushenko2020_CationPi_Inorganic}
These results are also supported by the relative peak intensity of the Na--N and Na--Bz at radial distribution functions (RDF, $g(r)$) (Figures S5 and S6).
Here, $g(r_{Na-N})$ and $g(r_{Na-Bz})$ provide clear structural fingerprints of sodium adsorption, in both cases the average-bond distance is 2.5 \AA~ (first peak in Figure \ref{fig:Fig_gr}A and S6). 
% -- Clusters
Furthermore, Hirshfeld charges reveal the sudden reduction in ionic character for some Na  (from $q\approx0.5\to0.17$), which, combined with $g(r_{Na-Na})$ in Figure~\ref{fig:Fig_gr}B, demonstrates the formation of sodium clusters~\cite{2026_NaClusters_Tateyama}. 
The $g(r_{Na-Na})$ for $x\gtrsim4$ is consistent with gas phase sodium clusters, with first neighbour bond distance 3.5 \AA{} (see Figure S4). 
This feature marks the formation of short-range Na--Na order and metallic-like behavior, as we will discuss next.

\begin{figure}[htp]
    \centering
    \includegraphics[width=1.0\columnwidth]{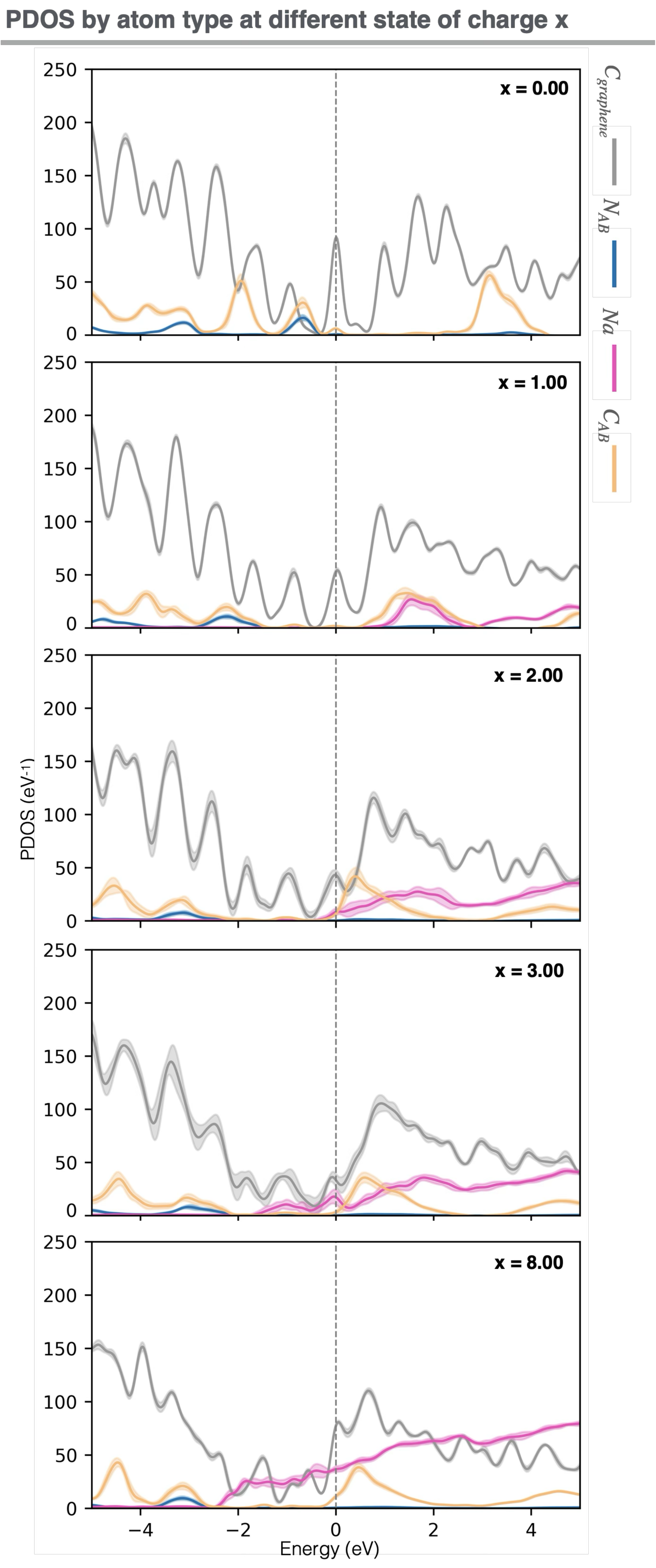}
    \caption{\textbf{Projected Density of States (PDOS) by atomic species in Na$_{x}$AB.} The PDOS is averaged over ten single-point calculations for each sodium concentration $x$. C$_\mathrm{AB}$, H$_\mathrm{AB}$, and N$_\mathrm{AB}$ denote atoms belonging to the aminobenzene groups. The Fermi level is set to 0 eV. Shaded regions indicate the standard deviation across sampled configurations.}
    \label{fig:pdos_type_x}
\end{figure}

\textit{Evolution of the Density of States with sodium content.} 
The density of states (DOS) is the electronic fingerprint of a material and provides insightful information on the evolution of the electronic states during electrode operation.
The electronic response of the anode to sodium insertion was examined through the projected DOS (PDOS) averaged over MD configurations, as shown in Figure \ref{fig:pdos_type_x}. 
At low Na content ($0<x<2$), the Fermi level lies within the states of the graphene layers with minor contributions from the aminobenzene groups. Sodium states lie above the Fermi level, consistent with their ionic Hirshfeld charges observed in this regime of concentration. 
As $x$ increases, the Na PDOS progressively populates states nearer the Fermi level, ultimately crossing it at intermediate sodium content ($x \approx 2$). This evolution reflects the gradual loss of ionic character and the emergence of partially delocalized, metallic-like Na states inside the material (Figure~\ref{fig:Fig_Average_H_Charges_vs_Na_x}). 
Meanwhile, the nitrogen states remain deep in the valence region across all sodium concentrations, confirming that \mbox{--NH$_{2}$} groups act primarily as electrostatic anchor sites rather than a significant participant in charge transfer. 
% -- hybridization
In contrast, the aminobenzene carbon states (i.e. $\pi$ states) increase in intensity near the Fermi level with increasing $x$, indicating hybridization between Na and the aromatic $\pi{}$ system.
An orbital-resolved PDOS and the charge-density difference ($\Delta{\rho{}}$) for a representative low-concentration configuration (Figure \ref{fig:delta_rho_x0.5}A at $x=0.5$) shows that Na $s$ and $p$ states lie within the same energy region as the C$_{AB}$ $p$ states corresponding to the aromatic $\pi$ ring, indicating Na$-\pi{}$ orbital hybridization: cation--$\pi$ bond~\cite{Dougherty2025_CationPi_ChemRev, Petrushenko2020_CationPi_Inorganic}. 
The $\Delta{\rho{}}$ isosurfaces reveal the characteristic behavior of this interaction, an extended accumulation of electron density across the benzene ring and into the adjacent graphene sheet (Figure \ref{fig:delta_rho_x0.5}B), accompanied by a localized depletion of electron density on the Na atom (Figure \ref{fig:delta_rho_x0.5}C). 
On the other hand, only minor density accumulation appears near the nitrogen of the amino group, confirming the predominantly electrostatic nature of the Na$^+\cdots$NH$_2-$ interaction.
A layer-resolved PDOS (Figure S3) reveals that this electronic behavior is consistent across the anode's layers.
These results further support the formation of sodium clusters and nanowires and the transition of sodium states to a metallic phase.

\textit{Voltage and Diffusion Trends across Sodium Concentration.} 
Among the quantities accessible from our simulations, the OCV profile is the most directly comparable to experiment and provides the key bridge between microscopic Na--storage dynamics and macroscopic battery performance.
The OCV in Figure~\ref{fig:AnodeChar}A) displays a high voltage stabilization region (up to $\sim 1.6$ V) for $x\lesssim1.5$, reflecting the large energetic stabilization associated with Na binding to the aminobenzene ring and amino sites.  
At these states of charge, Hirshfeld atomic charges remain highly positive, and Na-electronic states lie well above the Fermi level in the PDOS.
Structurally, the interlayer distance remains close to the pristine value ($\sim$9~\AA{}), providing an unhindered environment for Na atoms to migrate between adjacent high-affinity adsorption sites promoted by thermal excitations.
These structural and electronic features result in high diffusion coefficients ($\sim$3.5 × 10$^{-6}$~cm$^2$s$^{-1}$, Figure~\ref{fig:AnodeChar}C).
As sodium content increases ($1.5\lesssim x \lesssim 4$), the aminobenzene pillars begin to tilt due to the formation of highly energetically-stable Na$_4$@AB$_3$ pyramidal structure and fluxional Na$_3$@AB$_2$ triangular arrangements (see Animation 1), causing a drop in voltage below the reference sodium metal.
It is important to highlight that these trapped sodium clusters contribute to the irreversibility of the intercalation/deintercalation process (capacity loss) of the anode, consistent with the trapping motifs found in hard carbon~\cite{Reddy2018_SodiumInsertion_HC}. 
This reorganization produces a notable reduction of the interlayer spacing to $\sim$6.3~\AA{} (Figure \ref{fig:AnodeChar}B) accompanied by a sudden drop in ion diffusivity due to the strong ion confinement.
This ion concentration domain also coincides with the transition from a clear ionic behavior to cluster-formation and metallic state as revealed by the Hirshfeld analysis, the PDOS, and the Na--Na RDF.
In hard carbon, Tateyama \textit{et al.}~\cite{2026_NaClusters_Tateyama} attribute the slope-to-plateau diffusivity drop to Na accumulation and clustering in the \textbf{}pore/graphitic transition regions.
In Na$_x$AB, the OCV falling below 0 V vs Na/Na$^+$ ($x \approx 3$) indicates that additional Na insertion promotes Na$_n$@AB$_m$ structure formation -- equivalent to the sodium nucleation/plating-like regime reported in hard carbon ~\cite{Vizintin2025_NaNMR_HC}-- rather than gallery environment population in the Janus framework.

The material reaches ideal working conditions for $x\gtrsim4$, 
where sodium storage takes place in well-defined interlayer galleries and on aminobenzene spacer sites~\cite{Weaving2020_Sodiation_HardCarbon_Raman,Yu2025_HC_Na_clusters_AIMD,Reddy2018_SodiumInsertion_HC}, thereby producing an OCV plateau at $\approx\mathit{0.152}$ V.
Moreover, the already saturated AB-sites and the abundance of interlayer pathways, combined with the high clustering and atom-dissociation ratios of the sodium aggregates, facilitate Na atom mobility ($\sim$4.8$\times$10$^{-6}$ cm$^2$s$^{-1}$).
Even at unusually high states-of-charge ($x=8.5$), the large diffusivity values are maintained, enabled by the low-cohesive energy of the clusters (see Animation 2). These values are 2--3 orders of magnitude larger than the effective Na diffusivities reported for hard carbon electrodes~\cite{Eren2025_ThreeStageModel_HC} and comparable to reported in-plane Li diffusivities in graphite ($\sim10^{-7}$--$10^{-6}$~cm$^{2}$~s$^{-1}$)~\cite{Persson2010_LithiumDiffusion_Graphite}.
Overall, the material delivers capacity of $\sim400$~mAh g$^{-1}$ (Figure~\ref{fig:AnodeChar}A), surpassing the theoretical Li@graphite benchmark (372 mAh g$^{-1}$, LiC$_6$)~\cite{Asenbauer2020_Graphite_LIB_Review}.

\section{Conclusions}

To conclude, we have characterized at room temperature the OCV and mechanical properties of aminobenzene-functionalized Janus graphene, providing solid evidence that positions this material as a potential candidate for a high-energy and high-capacity sodium anode.
It sustains a working voltage of 0.15 V for a broad range of states-of-charge, reaching a prominent capacity of $\sim400$~mAh g$^{-1}$, constant cell volume, and large diffusivity under operating conditions.
Furthermore, at the atomistic level, the sodium insertion mechanism has three stages: adsorption on reactive sites (i.e. aminobenzenes), Na$_n$@AB$_m$ structure formation, and interlayer gallery filling, compared to the four stages in hard carbon.
As a key factor, the last stage enables constant-voltage values over a broad range of states-of-charge.
These findings not only showcase the great potential of functionalized Janus materials as electrodes for electrochemical energy storage but also demonstrate the paramount role of predictive simulations endowed by combining machine learning techniques and electronic structure calculations.
\section{Methods}
We combined first-principles calculations, machine-learned molecular dynamics at finite temperature and electronic structure analysis to characterize aminobenzene-functionalized Janus graphene. A reference database was generated from ab initio molecular dynamics calculations with the all-electron code FHI-aims, employing the Perdew-Burke-Ernzerhoff exchange-correlation functional with non-local many-body dispersion interactions. The data was used to train a SpookyNet potential with standard hyperparameters, which was subsequently employed to perform molecular dynamics simulations at 300 K in the NVT ensemble on a $4\times4\times4l$ hexagonal supercell with a 0.5 fs time step over 5 ns. 
To analyze the electronic origin of Na adsorption, representative snapshots from equilibrated trajectories at different Na concentrations were selected for DFT single point calculations on a $2\times2\times3l$ supercell. These calculations provided Hirshfeld charges, projected density of states, and charge density differences. Further methodological details are provided in the Methods section in the Supporting Information.

\section*{ACKNOWLEDGEMENTS}
H.E.S. acknowledges and expresses gratitude to Carlos Ernesto L\'opez Natar\'en for helping with the high-performance computing infrastructure and for his valuable support.
%\subsection*{Funding}
H.E.S. acknowledges support from CONACYT/SECIHTI-Mexico under Project CF-2023-I-468, DGTIC-UNAM under Project LANCAD-UNAM-DGTIC-419, and from DGAPA-UNAM under Projects PAPIIT No. IA106023 and IA105625.  
C.I.V. acknowledges support from SECIHTI via the Estancias Posdoctorales por México EPM(1) 2024 program, which funded the postdoctoral stay at the Instituto de Física.
K.R.M.\ was partly supported by the Institute of Information \& Communications Technology Planning \& Evaluation (IITP) grant funded by the Korea government (MSIT) (No. RS-2019-II190079, Artificial Intelligence Graduate School Program, Korea University) and grant funded by the Korea government (MSIT) (No. RS-2024-00457882, AI Research Hub Project).
O.T.U. contributed to this paper in an advisory capacity only.
Correspondence should be addressed to H.E.S. and K.R.M.\\

%\section*{Associated content}
\textbf{Data Availability Statement} \\
Additional data related to this paper that further support the findings of this study are available from the corresponding authors upon reasonable request.

%\subsection*{Competing interests.} The authors declare that they have no competing interests.

%%%%%%%%%%%%%%%%%%%%%%%%%%%%%%%%%%%%%%%%%%%%%%%%%%%%%%%%%%%%%%
%%%%%%%%%%% ==========================  APPENDIX  ========================== %%%%%%%%%
%%%%%%%%%%%%%%%%%%%%%%%%%%%%%%%%%%%%%%%%%%%%%%%%%%%%%%%%%%%%%%

\bibliography{references.bib}% Produces the bibliography via BibTeX.
%
%
%
%

% ****** End of file apssamp.tex ******

\end{document}